\documentclass{aastex}

\begin{document}

\title{Mass Loss Rates of Li-rich AGB/RGB Stars}
\shorttitle{Mass Loss Rates}
\shortauthors{Maciel and Costa}

\author{W. J. Maciel} \and \author{R. D. D. Costa}
\affil{Astronomy Department, University of S\~ao Paulo, Brazil}
\email{wjmaciel@iag.usp.br, roberto.costa@iag.usp.br}

\begin{abstract}
A sample of AGB/RGB stars with an excess of Li abundances is considered in order to estimate their mass loss rates. Our method is based on a correlation between the Li abundances and the stellar luminosity, using a modified version of Reimers formula. We have adopted a calibration on the basis of an empirical correlation between the mass loss rate and some stellar parameters. We conclude that most Li-rich stars have 
lower mass loss rates compared with the majority of AGB/RGB stars, which show no evidences of Li enhancements, so that the Li enrichment process is apparently not associated with an increased mass loss rate.
\end{abstract}


\section{Introduction}
\label{sec1}

It is well known that most metal-rich AGB/RGB stars present strong Li underabundances, in view of the fact that Li is easily destroyed in stellar interiors. However, several stars including Red Giant Branch (RGB) and Asymptotic Giant Branch (AGB) objects present some Li enrichment, which can be characterized by abundances $\epsilon$(Li) = 
$\log$ (Li/H) + 12 $>$ 1.5. The mechanism producing the Li excess is not clear, and possibilities include the Cameron-Fowler mechanism \citep{cameron}, presence of planets, etc. [see for example the recent discussions by \citet{casey} and \citet{kirby2}].

Li-enrichment has been associated with an enhanced mass loss ejection. \citet{ramiro1, ramiro2}, and \citet{monaco1} comment that some Li-rich giants show evidences of mass loss and chromospheric activity. However, \citet{fekel} and \citet{jasniewicz} suggested that no important mass loss phenomena are associated with these stars, which is supported by the results by \citet{lebzelter} based on $K - [12 \mu m]$ colours of the 3 Li-rich stars.  This is in agreement with a suggestion in the
literature [cf. \citet{mallik} and \citet{luck}] that an enhanced mass
loss would remove the stellar outer layers where most Li atoms are
located.

In order to clarify the possible association of higher mass loss rates with the Li enhancements in AGB/RGB giants, in this work we estimate the mass loss rates of a sample of Li-rich AGB/RGB stars based on a correlation between the Li-abundance and the stellar luminosity. We use a modified Reimers formula calibrated on the basis of an independently
derived empirical correlation between the mass loss rate and some stellar parameters as suggested in the literature. As a result, we estimate the mass loss rates of a large sample of AGB/RGB stars with well determined Li enhancements.

\section{The data}
\label{sec2}

The data for the Li-rich stars include RGB and AGB stars, and are based on the sample adopted in our previous papers \citep{mc2012,mc2015}, with additional data from \citet{mc2016} and \citet{casey}. The original sources of the data are: \citet{brown}, \citet{mallik}, \citet{gonzalez}, \citet{monaco1,monaco2}, \citet{kumar}, \citet{lebzelter}, 
\citet{kovari}, \citet{martell}, \citet{lyubimkov}, and \citet{casey}. Apart from their own data, the latter also gives a list of previously analyzed stars, mainly from \citet{ruchti} and \citet{kirby1,kirby2}. In order to apply our method, the effective temperature $T_{eff}$, gravity $\log g$, and Li abundance $\epsilon$(Li) of the Li-rich stars must be known. Applying this condition and removing some objects that lie 
outside the range of the adopted stellar parameters, a final sample of 159 Li-rich stars is obtained. \footnote{A detailed table with the full list of objects, input data, and mass loss rates can be obtained from the corresponding author}

\section{The $\epsilon$(Li) $\times$ $\log L/L_\odot$ correlation}
\label{sec3}

From \citet{mc2016} we obtain a plot of the Li abundances as a function of the luminosities for a selected sample of Li-rich stars containing 57 objects in the range $0 < \log L/L_\odot < 2.6$, as shown in Figure~1 (empty circles). It can be seen that the Li-enhancements show some dispersion for each selected luminosity, since for some stars Li may have been more strongly destroyed than for others. In other words, an upper envelope can be observed in the maximum Li abundances, which represents the maximum Li enrichment at each luminosity interval.  The maximum Li enrichment presents a clear dependence on luminosity, in  the sense that the most luminous giants reach larger Li abundances. Therefore, we will choose the {\it maximum} contribution at each bin as representative of the Li enhancement process. Adopting 9 luminosity bins, we get the results shown by the black circles, where the error bars show the average dispersion at each bin.

\begin{figure}
\centerline{\includegraphics[angle=0, width=12.0cm]{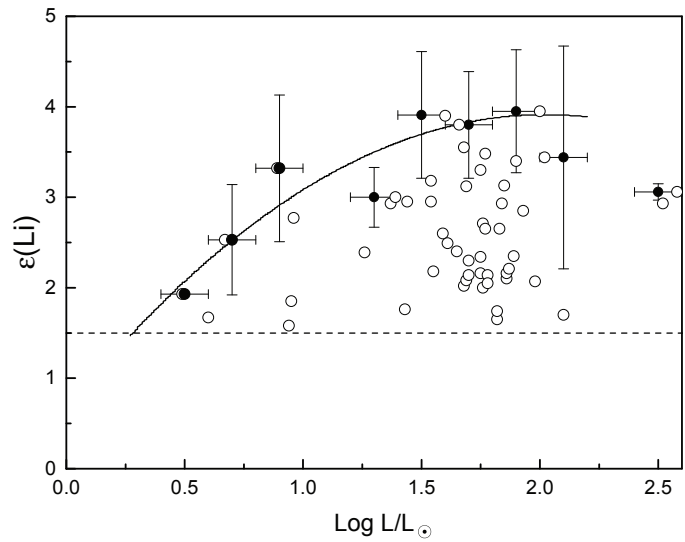}}
\caption[]{Li abundances as a function of the luminosity for Li-rich 
stars. Empty circles: data for stars with well-determined abundances and luminosities; filled circles: maximum abundances in each luminosity bin; the error bars show  the average dispersion at each bin; solid line: 
quadratic fit of the maximum abundances; dashed line: adopted baseline 
value for Li-rich stars, $\epsilon$(Li) = 1.5.}
\end{figure}

It is clear that the maximum Li abundance increases with the stellar luminosity up to $\log L/L_\odot \simeq  2.2$, but the behaviour of the correlation is not clear for the high luminosity stars. Since the  vast majority of stars in our sample have luminosities similar or lower than this limit (49 stars out of 57, or 86\%), we will adopt the following 
ranges where the correlation is better determined, that is

\begin{equation}
1.5 \leq \epsilon({\rm Li}) \leq 4.0  \ \ \ \ \ \ \ 0.2 \leq \log L/L_\odot \leq 2.2
\label{eq1}
\end{equation}

In Figure~1 the dashed line shows the baseline corresponding to the limit of Li-rich stars, for which $\epsilon$(Li) $\geq 1.5$, and the solid line is a polynomial fit to the maximum abundances at each luminosity bin given by

\begin{equation}
\epsilon({\rm Li}) = a + b \ \log L/L_\odot + c \ (\log L/L_\odot)^2 
\label{eq2}
\end{equation}

\noindent
where $a = 0.657 \pm 0.937$, $b = 3.221 \pm 1.771$ and  $c = -0.797 \pm 0.734$. This equation is assumed to be valid in the intervals given by Equation~(1), and can be easily solved for the luminosity.

Since the observed Li abundance may have any value lower or equal to the maximum value, it can be seen that the corresponding luminosity calculated by the solution of Equation~(2) is generally a lower limit. For lower values of $\epsilon$(Li), close to the minimum value of 1.5, the uncertainty is larger, since the stellar luminosity can be much larger than the value obtained from Equation~(2). On the other hand, for the values of the Li abundances close to the maximum value of 4.0 the luminosity is better determined, since lower luminosities are excluded from the maximum values adopted in each bin. Since it is assumed 
that the correlation is valid for luminosities up to $\log L/L_\odot \leq 2.2$, we have $\log L/L_\odot \leq \log L/L_\odot {\rm (true)} \leq 2.2$. Naturally, this excludes the objects with luminosities higher than 2.2, which are a small fraction of the Li-rich objects, in agreement with the data in Figure~1.

\section{Mass Loss rates of Li-rich AGB/RGB Stars}
\label{sec4}

In order to determine the mass loss rate we have adopted the following procedure: for each Li-rich star we consider the Li abundance $\epsilon$(Li) and estimate the luminosity using Equation~(2). From the luminosity and the effective temperature, the stellar mass can be estimated using recent detailed evolutionary tracks for giant stars. We have adopted the tracks by \citet{bertelli}, see also \citet{kumar}. The tracks can be 
applied to solar metallicity stars with masses in the interval $1.0 <M/M_\odot < 3.0$, and effective temperatures in the approximate range $3800 < T_{eff} (K) < 5600$. From the effective temperature and luminosities, the determination of the stellar mass is a straighforward procedure.  Using the stellar gravity $g$ taken from the same sources  as the effective temperature and Li abundances, the radius can be simply estimated by $R^2 = G \, M / g$. 

In order to obtain the mass loss rates (in $M_\odot$/yr), we have 
adopted a modified version of Reimers formula given by

\begin{equation}
{dM \over dt} = 4 \times 10^{-13} \, \eta \ {(L/L_\odot)\,(R/R_\odot) \over (M/M_\odot)}
\label{eq3}
\end{equation}

\noindent
[see for example  \citet{lamers}]. The $\eta$ parameter is considered as a free parameter, to be determined on the basis of an adequate calibration involving independently derived mass loss rates of AGB/RGB stars. From Equation~3 it can be observed that the mass loss rates increase as the luminosity increases. This is a result that was already obtained earlier in the literature [see for example \citet{van loon}], and is probably related to the fact that the mass loss is caused by the action of the stellar radiation pressure on grains, atoms and ions in the stellar
atmosphere. This can be seen in Figure~3,  where the solid dots represent Li-rich stars, and a particular value of parameter $\eta$ was used, as we will discuss in Section~5. Of course, the mass loss phenomenon 
may also be affected by other parameters, such as the chemical
composition, ionization state, etc., but it seems clear that
the main mechanism is related to the stellar radiation, so that a 
relation between the mass loss rate and luminosity is expected.

\citet{van loon} derived an empirical formula to estimate the mass loss rate for oxygen-rich AGB stars and red supergiants as a function of the stellar luminosity and effective temperature. The formula is based on the modelling of the spectral energy distributions of a sample of red giants in the Large Magellanic Cloud. It is believed that the mass loss process in these stars originates from the action of the stellar radiation pressure on solid grains in the external stellar layers, so that it is expected that the mass loss rate depends on the stellar luminosity,
responsible for the radiation pressure, as well as the stellar temperature, which affects  the process of grain formation. Previous results by \citet{van loon2} have already shown that the measured mass loss rates in LMC AGB stars generally present an increase with the stellar luminosity, which is in agreement with our adopted modified Reimers formula. In \citet{van loon} the mass loss rates were derived using a dust radiative transfer code applied to infrared photometry data to obtain the spectral energy distribution. The derived equation  can be written as

\begin{equation}
\log {dM\over dt} = \alpha + \beta \ \log\biggl({L \over 10000\, L_\odot}\biggr) + \gamma \ \log\biggl({T_{eff} \over 3500\, {\rm K}} \biggr)
\label{eq4}
\end{equation}

\noindent
where the mass loss rates are given in $M_\odot/{\rm yr}$.
For M-type stars the constants are $\alpha = -5.64 \pm 0.15$, $\beta = 1.05 \pm 0.14$ and $\gamma = -6.3 \pm 1.2$. This corresponds to an approximately linear relation between the mass loss rate and the stellar luminosity, in agreement with predictions from dust radiative driven winds. From Equation~4 it can also be seen that the mass loss rate decreases with increasing temperatures, which is expected on the basis of the dust formation process. The uncertainty in the mass loss rates from Equation~4 is estimated as $\Delta \log dM/dt \simeq 0.3$, or approximately a factor two for typical mass loss rates of $10^{-5}\, M_\odot/$yr expected for the most luminous objects. 

Although based on LMC objects which typically have lower metallicities than  Galactic objects, Equation~4 can be applied to Galactic objects as well, as shown by a comparison of the mass loss rates in the LMC with indepently derived rates for galactic AGB stars, as discussed by \citet{van loon}. In other words, the mass loss rates are not strongly affected by the metallicity, at least within the estimated uncertainties.

\begin{figure}
\centerline{\includegraphics[angle=0, width=12.0cm]{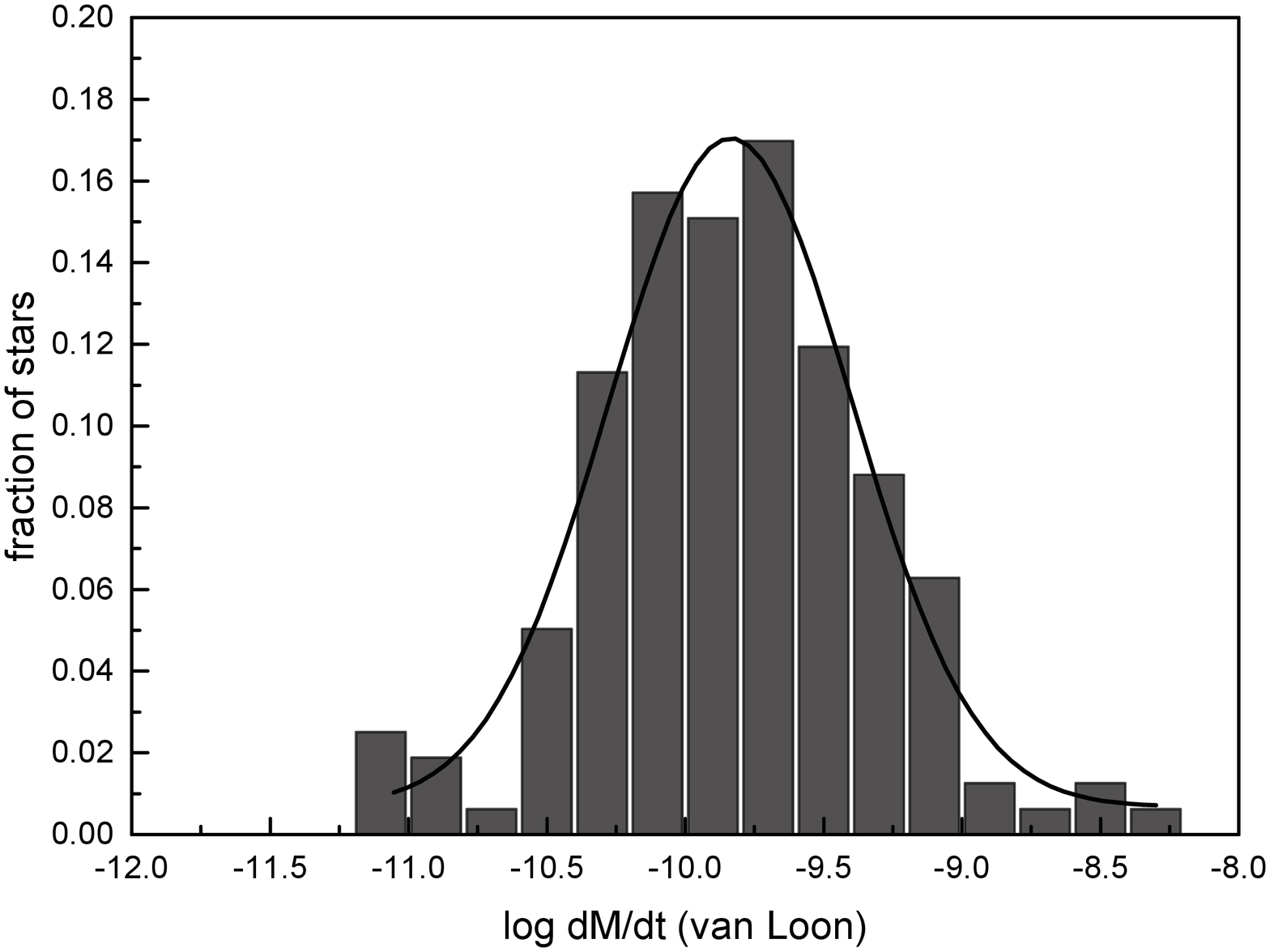}}
\end{figure}

\begin{figure}
\centerline{\includegraphics[angle=0, width=12.0cm]{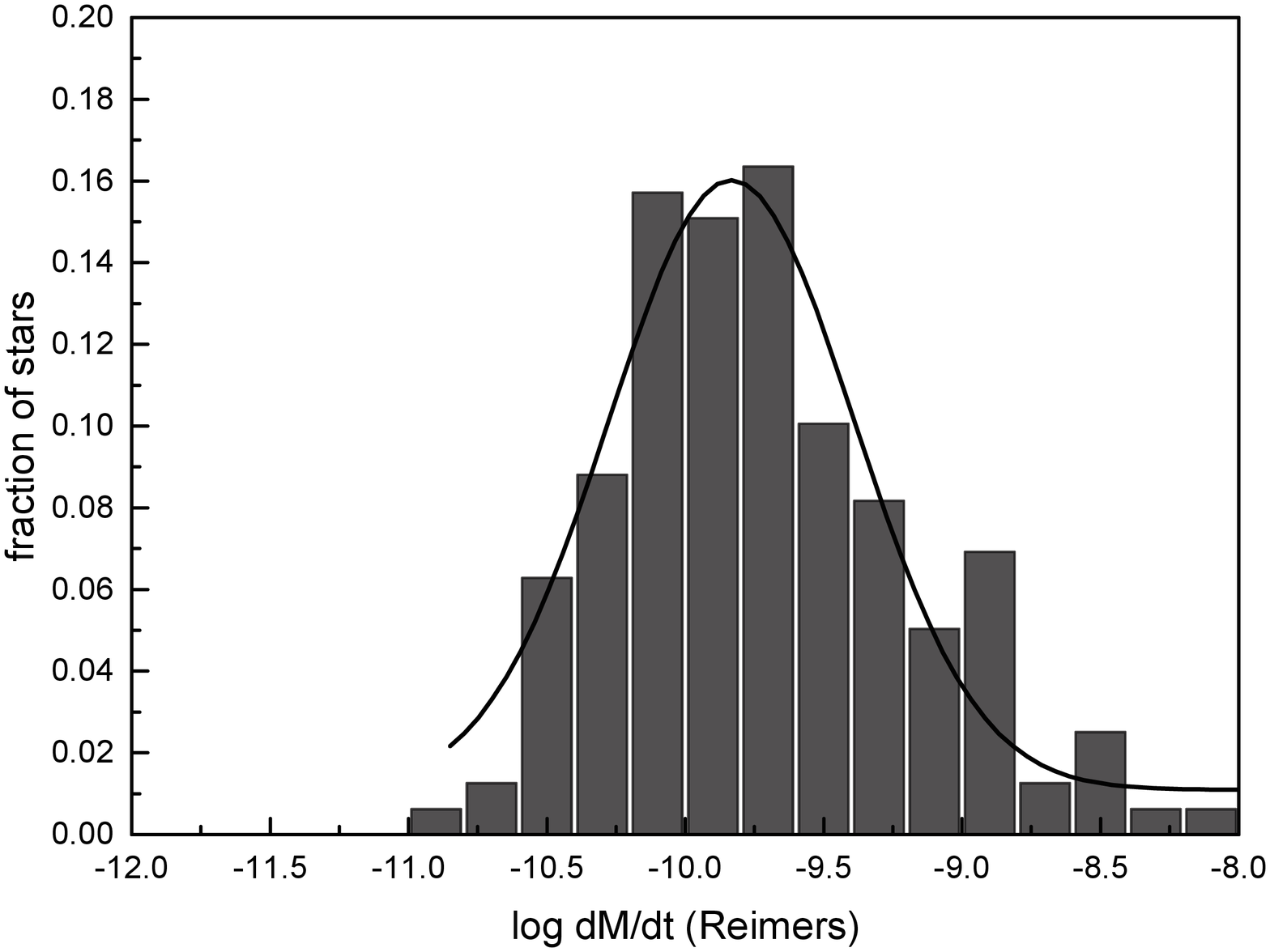}}
\caption[]{Distribution of the mass loss rates of Li-rich stars using (a) the empirical formula by van Loon, and (b) the results of the 
modified Reimers formula for $\eta = 5.7$. The solid  curves are gaussian fits.}
\end{figure}

\section{Results and discussion}
\label{sec5}

We have applied Equation~4 to our sample of Li-rich AGB stars adopting the luminosities derived from the correlation between the stellar luminosity and the Li-abundance. As a result we obtain the distribution shown in Figure~2a, where the solid curve represents a gaussian distribution. In order to calibrate our method, we have adopted a linear relation between the mass loss rate and the luminosity of the form $\log dM/dT =  A + B   \log L/L_\odot$. It is easy to see that the slope of this  relation 
does not depend on the $\eta$ parameter. Applying this correlation to the sample of Li-rich stars with 159 objects, we have $B =  1.057 \pm 0.105$.
The intercept $A$ can be obtained provided we estimate the $\eta$ parameter on the basis of an adequate calibration. This can be achieved by selecting the $\eta$ value that reproduces the distribution given by Figure~2a. As a result we have $\eta = 5.7$ with $A = -10.620\pm 0.099$ and $B = 1.057\pm 0.105$, with a correlation coefficient 
$r = 0.63\pm 0.45$. The corresponding distribution is shown in Figure~2b, and it can be seen  that both distributions are very similar.

The uncertainties of the derived mass loss rates can be roughly estimated by considering that typical uncertainties in the Li abundances are of the order of 0.20 dex; for the effective temperature we have uncertainties better than 100 K for most stars; the stellar gravity has a typical uncertainty of 0.20 dex. From the adopted correlation involving the 
 stellar luminosity an average uncertainty of about 0.20 dex is expected. Therefore, we have an uncertainty of about $0.5\,M_\odot$ for the stellar mass, and a final uncertainty of about 0.50 dex for the mass loss rate $\log dM/dt$. This is comparable with the uncertainties in the mass loss rates of AGB/RGB stars with no indications of Li-enhancements, as given
by \citet{gullieuszik}. In this case, an average dispersion of about 0.5 dex for $\log dM/dt$ corresponds roughly to a factor 2 for a typical mass loss rate of $ dM/dt \sim 10^{-6}\, M_\odot$/year. \citet{groenewegen} give a slightly smaller uncertainty of 0.43 dex in for AGB stars and red supergiants in the Magellanic Clouds. Since our main goal is to compare 
the mass loss rates of the Li-rich stars and the Li-poor stars, the absolute values of the rates are of secondary importance.

Figure~3 shows the derived mass loss rates as a function of the luminosity for our full stellar sample, containing 159 stars. In this figure the black dots on the left side of the figure are the results using the modified Reimers formula (Equation~3), while the crosses indicate the rates obtained by the empirical formula by van Loon (Equation~4). The dashed line shows the final correlation obtained for Li-rich stars. It can be seen that the adopted linear correlation between the mass loss rates and the luminosities is a realistic one in the adopted luminosity range given by Equation~1.  We would like to stress that the dashed line is not a fit for the Li-poor stars, as the corresponding sample is far from complete. We  have selected a significant number of objects to stress the apparent dichotomy between Li-poor and Li-rich stars, which is a reflection of the lower luminosities of the latter.

\begin{figure}
\centerline{\includegraphics[angle=-90, width=12.0cm]{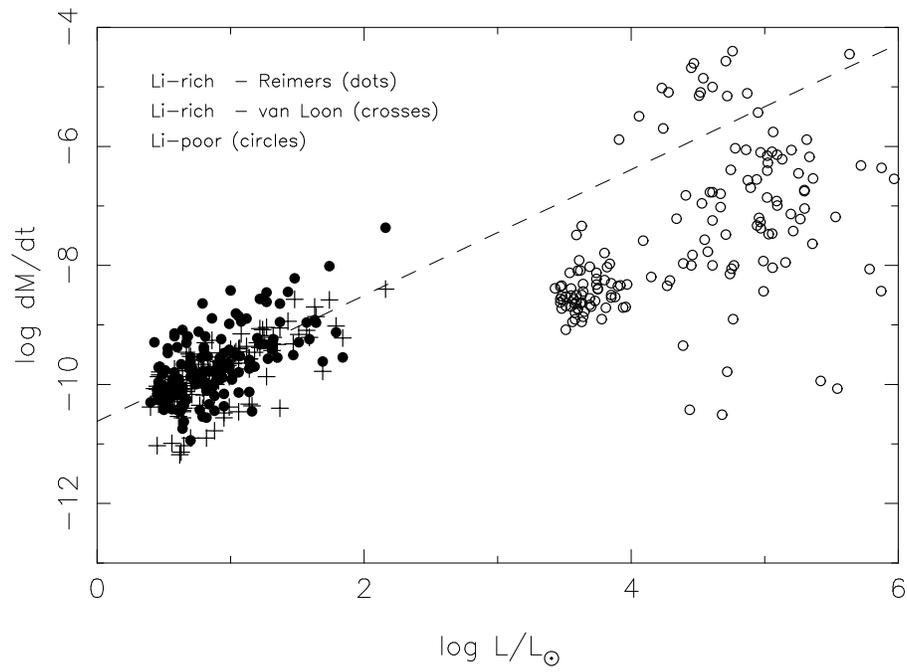}}
\caption[]{Mass loss rates ($M_\odot/{\rm yr}$) as a function of 
luminosity. Dots: Li-rich stars, Modified Reimers Formula; crosses: 
Li-rich stars, van Loon Equation; empty circles:  Li-poor stars from the literature; dashed line: linear correlation for the Li-rich stars.}
\end{figure}

As we have mentioned before, in view of the adopted correlation between the Li abundance and the luminosity, it can be seen from Figure~1 that the objects with lower values of $\epsilon$(Li) may have higher luminosities than those obtained by Equation~(2). For intermediate values of $\epsilon$(Li) the uncertainty is lower, and for the highest values of the Li abundance the calculated values are approximately correct, since the derived luminosities are close to the maximum value. This means that our luminosities should be considered as lower limits, especially for those stars with Li abundances $\epsilon$(Li) $< 2.0$, approximately, which are a small fraction of the objects in our sample (about 8\%). As a consequience, the position of the stars on the $dM/dt \times L/L_\odot$ plane is more strongly affected for the objects with lower Li abundances.  This conclusion has been confirmed by simulations for a few stars with different Li enhancements within the range shown in Figure~1. As expected, the objects with lower enhancements are slightly displaced upwards in Figure~3, while those stars near the maximum Li abundances remain essentially at the same position in the diagram, so that the general trend  shown by the solid dots of Figure~3 is not significantly changed. 

There are many reliable determinations of the luminosities and mass loss rates of AGB/RGB stars with no evidences of Li enhancements in the literature. As an example, we have considered the samples by \citet{gullieuszik} and \citet{groenewegen}, selecting the O-rich stars. We have also included the results for Local Group galaxies by
\citet{gs2018} again selecting the O-rich objects. Excluding objects for which a complete calculation could not be made due to the lack of accurate determinations of the mass loss rate $dM/dt$ and/or the luminosity $\log L/L_{\odot}$, we have a final sample of 156 stars, which is a representative set for these objects. The estimated uncertainties are generally of 10\% for the luminosity and 25\% in the mass loss rate. These objects are also included in Figure~3 as empty circles, mostly located on the right side of the figure. It is possible that some of the luminous stars at the right side of Figure~3 may have some excess Li, but it should be recalled  that the Li-rich stars are a tiny fraction of all RGB/AGB stars, so that our assumption that most of the objects on the right side of the figure are indeed Li-poor is quite reasonable. It can be seen that most of these objects have higher luminosities and mass loss rates compared with the Li-rich stars, with very few exceptions. It can then be concluded that the results obtained by the modified Reimers formula indicate that the Li-rich objects are generally associated with mass loss rates much lower than in the case of the majority of AGB/RGB stars, which are Li-poor objects.  In other words, Li enhancements seem to be a low-luminosity feature associated with lower mass loss rates compared with the majority of these stars, in agreement with our preliminary estimates using a linear correlation between the Li abundances and the luminosities, as discussed in \citet{mc2016}.

\acknowledgements

We are indebted to Dr. J. van Loon for some very interesting comments on an earlier version of this paper. This work was partially supported by CNPq (Process 302556/2015-0) and FAPESP (Process 2010/18835-3).

%

\bibliographystyle{an}
\bibliography{}

\begin{thebibliography}


%

\bibitem[Bertelli et al. (2008)]{bertelli}
Bertelli, G., Girardi, L., Marigo, P., Nasi, E. 2008, \aap, 484, 815

\bibitem[Brown et al. (1989)]{brown}
Brown, J. A., Sneden, C., Lambert, D. L., Dutchover, E. 1989, \apjs, 71, 293

\bibitem[Cameron \& Fowler (1971)]{cameron}
Cameron, A. G. W., Fowler, W. A. 1971, \apj, 164, 111

\bibitem[Casey et al. (2016)]{casey}
Casey, A. R., Ruchti, G., Masseron, T., et al. 2016, \mnras, 461, 3336

\bibitem[de La Reza et al. (1996)]{ramiro1}
de La Reza, R., Drake, N. A., da Silva, L. 1996, \apj, 456, L115

\bibitem[de La Reza et al. (1997)]{ramiro2}
de La Reza, R., Drake, N. A., da Silva, L., Torres, C. A. O., Martin, E. L. 1997, \apj,  482, L77

\bibitem[Fekel \& Watson (1998)]{fekel}
Fekel, F. C., Watson, L. C. 1998, \aj,  116, 2466

\bibitem[Gonzalez et al. (2009)]{gonzalez}
Gonzalez, O. A., Zoccali, M., Monaco, L., Hill, V., Cassisi, S., Minniti, D., Renzini, A., Barbuy, B., Ortolani, S., Gomez, A. 2009, \aap,  508, 289

\bibitem[Groenewegen et al. (2009)]{groenewegen}
Groenewegen, M. A. T., Sloan, G. C., Soszy\'nski, I., Peterson, E. A. 2009, \aap,  506, 1277

\bibitem[Groenewegen \& Sloan (2018)]{gs2018}
Groenewegen, M. A. T., Sloan, G. C. 2018, \aap, 609, A114

\bibitem[Gullieuszik et al. (2012)]{gullieuszik}
Gullieuszik, M., Groenewegen, M. A. T., Cioni, M. R. L., de Grijs, R., van Loon, Th., Girardi, L., Ivanov, V. D., Oliveira, J. M., Emerson, J. P., Goardalini, R. 2012, \aap, 537, A105

\bibitem[Jasniewicz et al. (1999)]{jasniewicz}
Jasniewicz, G., Parthasarathy, M., de Laverny, P., Th\'evenin, F. 1999, \aap, 342, 831

\bibitem[Kirby et al. (2012)]{kirby1}
Kirby, E. N., Fu, X., Guhathakurta, P., Deng, L. 2012, \apjl, 752, L16

\bibitem[Kirby et al. (2016)]{kirby2}
Kirby, E. N., Guhathakurta, P., Zhang, A. J., et al. 2016, \apj, 819, 135

\bibitem[K\"ov\'ari et al. (2013)]{kovari}
K\"ov\'ari, Zs., Korhonen, H., Strassmeier, K. G., Weber, M., Kriskovics, L., Savanov, I. 2013, \aap,  551, A2

\bibitem[Kumar et al. (2011)]{kumar}
Kumar, Y. B., Reddy, B. E., Lambert, D. L. 2011, \apj,  70, L12

\bibitem[Lamers \& Cassinelli (1999)]{lamers}
Lamers, H. J. G. L., Cassinelli, J. 1999, Introduction to stellar
winds, Cambridge

\bibitem[Lebzelter et al. (2012)]{lebzelter}
Lebzelter, T., Uttenthaler, S., Busso, M., Schultheis, M., Aringer, B. 2012, \aap, 538, A36

\bibitem[Luck (1977)]{luck}
Luck, R. E. 1977, \apj, 218, 752

\bibitem[Lyubimkov et al. (2012)]{lyubimkov}
Lyubimkov, L. S., Lambert, D. L., Kaminsky, B. M., Pavlenko, Y. V., Pokland, D. B., Rachkovskaya, T. 2012, \mnras, 427, 11

\bibitem[Maciel \& Costa (2012)]{mc2012}
Maciel, W. J., Costa, R. D. D. 2012, \memsai, 22, 103

\bibitem[Maciel \& Costa (2015)]{mc2015}
Maciel, W. J., Costa, R. D. D. 2015, Why galaxies care about AGB stars III, ASP CS 497, 313

\bibitem[Maciel \& Costa (2016)]{mc2016}
Maciel, W. J., Costa, R. D. D. 2016, The 19th Cambridge Workshop on Cool Stars, Stellar Systems, 
and the Sun, ed. G. A. Feiden, https://zenodo.org/record/59278\#.V6iCq6JwspI

\bibitem[Mallik (1999)]{mallik}
Mallik, S. V. 1999, \aap, 352, 495

\bibitem[Martell \& Shetrone (2013)]{martell}
Martell, S. L., Shetrone, M. D. 2013, \mnras, 430, 611

\bibitem[Monaco et al. (2011)]{monaco1}
Monaco, L., Villanova, S., Moni Bidin, C., Carraro, G., Geisler, D., Bonifacio, P., Gonzalez, O. A., Zoccali, M., Jilkova, L.  2011, \aap, 529, A90
 
\bibitem[Monaco et al. (2014)]{monaco2}
Monaco, L., Boffin, H. M. J., Bonifacio, P., Villanova, S., Carraro, G., Caffau, E., Steffen, M., Ahumada, J. A., Beletsky, Y., Beccari, G.  2014, \aap, 564, L6

\bibitem[Ruchti et al. (2011)]{ruchti}
Ruchti, G. R., Fullbright, J. P.,  Wyse, R. F. G., et al. 2011, \apj, 743, 107

\bibitem[van Loon (2000)]{van loon2}
van Loon, J. Th. 2000, \aap, 354, 125
 
\bibitem[van Loon et al. (2005)]{van loon}
van Loon, J. Th., Cioni, M. R. L., Zijlstra, A. A., Loup, C. 2005, \aap, 438, 273

\end{thebibliography}

\end{document}